\documentclass[conference]{IEEEtran}
\makeatletter
\def\ps@headings{%
\def\@oddhead{\mbox{}\scriptsize\rightmark \hfil \thepage}%
\def\@evenhead{\scriptsize\thepage \hfil \leftmark\mbox{}}%
\def\@oddfoot{}%
\def\@evenfoot{}}
\makeatother
\usepackage{booktabs}
\usepackage{amssymb}
\usepackage{amsmath}
\usepackage{graphicx}
\usepackage[caption=false,font=footnotesize]{subfig}
\usepackage{balance}

\newcommand{\mfrac}[2]{\ensuremath{%
    \frac{\raisebox{-.2ex}{\ensuremath{#1}}}%
         {\raisebox{.2ex}{\ensuremath{#2}}}}}  

\newcommand{\ra}[1]{\renewcommand{\arraystretch}{#1}}

\newtheorem{theorem}{Theorem}[section]
\newtheorem{lemma}[theorem]{Lemma}

\begin{document}
%\crdata{978-1-60558-334-1/08/10} 

\title{Multiple Random Walks to Uncover \\ Short Paths in Power Law Networks}

\author{
	UMass CMPSCI TechReport UM-CS-2011-049 \\
	~\\
 	\begin{tabular}{cc}
 		\multicolumn{2}{c}{Bruno Ribeiro$^{1}$, Prithwish Basu$^2$, and Don Towsley$^{1}$} \tabularnewline
 		&
 		\tabularnewline
 		$^{1}$Computer Science Department & $^{2}$Raytheon BBN Technologies
 		\tabularnewline
 		University of Massachusetts Amherst & Cambridge, MA
 		\tabularnewline
 		\{ribeiro, towsley\}@cs.umass.edu & pbasu@bbn.com 
 	\end{tabular}
}

\maketitle

\begin{abstract}
Consider the following routing problem in the context of a large scale network $G$, with particular interest paid to power law networks, although our results do not assume a particular degree distribution.
A small number of nodes want to exchange messages and are looking for short paths on $G$. 
These nodes do not have access to the topology of $G$ but are allowed to crawl the network within a limited budget.
Only crawlers whose sample paths cross are allowed to exchange topological information.
In this work we study the use of random walks (RWs) to crawl $G$.
We show that the ability of RWs to find short paths bears no relation to the paths that they take.
Instead, it relies on two properties of RWs on power law networks:
\begin{enumerate}
\item RW's ability observe a sizable fraction of the network edges; and
\item an almost certainty that two distinct RW sample paths cross after a small percentage of the nodes have been visited.
\end{enumerate}
We show promising simulation results on several real world networks.
\end{abstract}

\section{Introduction}

Consider the following routing problem in the context of a large scale network $G$ with $n \gg 1$ nodes, with particular interest paid to power law networks, although our results do not assume a particular degree distribution.
A small set of nodes ($h$ in total) want to exchange messages and are looking for short paths on $G$.
We consider routing in two phases: Topology discovery and shortest paths calculation.
In the topology discovery phase each node must crawl the network in order to partially discover the network topology.
Nodes have limited  crawling budget and only crawlers whose sample paths intersect are allowed to exchange topological information.
Upon finishing the topology discovery phase nodes are allowed to optimally route on the sampled topology.

In this work consider crawling $G$ with $h$ independent random walkers (RWs).
We show that the ability of RWs to find short paths bears no relation to the path that they take.
Instead, it relies on two RW properties on power law networks:
\begin{itemize}
\item RW's ability observe a sizable fraction of the network edges; and
\item two RW sample paths (starting at different nodes) cross after a small percentage of the nodes have been visited, with high probability.
\end{itemize}
We provide promissing simulations results on several real world networks.

\begin{figure}
\centering
\includegraphics[angle=90,scale=0.25,type=pdf,ext=.pdf,read=.pdf]{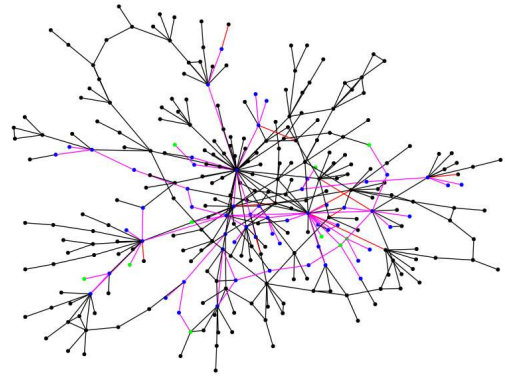}
\vspace{-10pt}
\caption{Illustration of a RW sample path. Green nodes are the starting nodes, the blue nodes are nodes visited by the random walkers, and the purple edges are the edges used by the walkers to explore the graph which is a subset of $E(B,i)$.\label{fig:progress}}
\vspace{-10pt}
\end{figure}

\subsection{Framework}
In what follows we formalize the model.
Consider an undirected graph $G=(V,E)$ with $n$ nodes and $m$ edges.
Start $h$ random walkers from nodes $U = \{u_1,\dots,u_h\}$ to explore the network.
We assume each walker takes $B$ steps. 
Each walker leaves bread crumbs so that one can trace its path back towards the initial node.
Let $X(t,i)=(x_1,\dots,x_t)$, $i=1,\dots,h$, denote the sequence of nodes seen by the $i$-th walker at step $t$.
Let $S(t,i)$, $i=1,\dots,h$, denote the set of nodes visited by the $i$-th walker at time $t$, i.e., $S(t,i)$ is $X(t,i)$ without repeated nodes.
And let $E(B,i)$ denote the set of edges belonging to the nodes visited by the $i$-th walker, $S(B,i)$, after the walker budget $B$ is exhausted.
Upon exausting its budget RW $i$ outputs $X(B,i)$, $S(B,i)$, and $E(B,i)$.

\subsection{Naive Routing}
Note that if $S(B,i) \cap S(B,j) \neq \emptyset$, $i,j=1,\dots,h$, then walker $i$ must have found walker $j$'s breadcrumbs or vice-versa.
Without loss of generality assume walker $i$ found $j$'s breadcrumbs.
A naive way to construct a path between $u_i$ and $u_j$ is to route messages retracing all RW steps.
Node $u_i$ acomplishes this by retracing $X(B,i)$ to contact one of the nodes in $v \in S(B,i) \cap S(B,j)$. 
Once the message reaches node $v$, it reaches its final destination, node $u_j$, following $j$'s breadcrumbs starting from $v$, i.e., the message follows the reverse RW path from $u_j$ to $v$.
The breadcrumb is installed only at the first visit to a node, which guarantees that the reverse RW path is without loops.

\subsection{Drawbacks of Naive Routing}
The naive algorithm presented above is quite inefficient.
Random walks are not particularly good at finding short paths (as seen in Section~\ref{sec:discussion}); unless there are few paths between pairs of nodes, e.g., $G$ is a tree.
A random walk with erased loops, also known as a loop-erased random walk,  samples all possible spanning trees of $G$ with equal probability~\cite{looperased}.
This can lead to long paths lengths between $u_i$ and $u_j$.

\subsection{The True Power of RWs in Finding Short Paths}
The ability of RWs to find short paths bears no relation to the paths that they take.
Instead, it relies on RW properties found under conditions often present in power law networks.
We summarize the main results of Section~\ref{sec:RWprop} here:
\begin{itemize}
\item $\vert E(B,i) \vert \approx  \frac{\langle k^2 \rangle - \langle k\rangle}{\langle k \rangle} B$, where $B \ll n$, $i=1,\dots,h$, and $\langle k \rangle$ and $\langle k^2 \rangle$ are the first and second moment of the node degrees in original graph $G$, respectively.
\item Under mild conditions two RW sample paths cross, with high probability.
\end{itemize}
These observations motivate our use of RWs to partially uncover the topology of $G$.

\begin{table}
\centering
\ra{1.2}
\begin{tabular}{@{}lllll@{}}
\toprule
\vspace{4pt}
Network & nodes & edges & $\langle k \rangle$ &  $\frac{\langle k^2 \rangle - \langle k\rangle}{\langle k \rangle}$ \\
\hspace{2pt}
Flickr~\cite{Mislove}  & 1.7M &  31.1M & 18.1 & 943.4 \\
\hspace{2pt}
AS Net.~\cite{CAIDA}  & 26.5K & 107K & 4 & 279.24 \\
\hspace{2pt}
Livejournal~\cite{Mislove} & 5.2M & 9.8M & 18.9 & 154 \\
\hspace{2pt}
Enron~\cite{Enron} & 37K & 368K & 10 & 140 \\
\hspace{2pt}
Gnutella~\cite{Gnutella} & 62.5K & 296K & 4.7 & 10.6 \\
\hspace{2pt}
Power Grid~\cite{PowerGrid} & 5K & 13K & 2.7 & 2.9 \\
\bottomrule
\end{tabular}
\caption{Summary of networks\label{tab:traces}}
\vspace{-25pt}
\end{table}

\subsection{Outline}
The outline of this work is as follows.
Section~\ref{sec:RWprop} characterizes the subgraph obtained by the $i$-th random walker and the probability that two RW sample paths cross.
Section~\ref{sec:algo} presents the Random Walk Short Path (RWSP) algorithm.
Section~\ref{sec:results} presents our results when simulating RWSP on real world networks.
Finally, Section~\ref{sec:discussion} presents a discussion and related work.

%\begin{itemize}
%\item Is there a systematic, analytical characterization of the shortest path lengths of a subgraph obtained by random walks?
%\item How does the histogram vary as a function of k and B and the structural properties of G?
%\end{itemize}

\section{RW Subgraph Properties} \label{sec:RWprop}

In this section we show that the real power of a random walk resides in its ability to collect edges from $G$.
Let $E(B,i)$ denote the set of edges belonging to the nodes visited by the $i$-th walker, $S(B,i)$, after the walker budget $B$ is exhausted.
We see that RW $i$, $i=1,\dots,h$, builds an edge rich subgraph $G^\prime(B,i)=(V,E(B,\cdot))$ with budget $B = \beta n$ in a network with large degree variance.
In what follows we assume $n \gg 1$, $B \ll n$, and that RW $i$ has access to the neighbor list of each visited node, such that $E(B,i)$ is obtained without extra sampling costs.

These are the main results of this section:
\begin{itemize}
\item $\vert S(B,\cdot) \vert \approx B$, i.e., the set of nodes visited by the RW scales linearly with $B$. The RW tends to collect only new nodes at the beginning of the walk.
\item $\vert E(B,i)\vert \approx  \frac{\langle k^2 \rangle - \langle k\rangle}{\langle k \rangle} B$, where $\langle k \rangle$ and $\langle k^2 \rangle$ are the first and second moment of the node degrees in original graph $G$, respectively.
\item We provide conditions under which there is a RWSP path between any pair of nodes $(u_i,u_j)$, $\forall i, j$, with high probability.
\end{itemize}
In power law networks $q = ({\langle k^2 \rangle - \langle k\rangle})/{\langle k \rangle}$ is likely to be quite large, and thus $G^\prime$ spans a sizable fraction of $G$ edges even when $B$ is a small fraction of the total number of nodes.
For instance, in the Flickr social photo sharing website $q = 943.4$, in the Enron email network this value is $q=139$, and in the Livejournal blog posting social network this value is $q=154$. 
Table~\ref{tab:traces} presents values of $q$ for several real world networks.

\subsection{The RW subgraph $G^\prime(B,i)$ contains a sizable fraction of the edges of $G$. \label{sec:ne}}
In what follows we find a closed approximation formula that gives the number of edges in $G^\prime(t,i)$, $t \leq B$.
Let $S(t,i)$ denote the set of nodes visited by the $i$-th walker at time step $t$, $i=1,\dots,h$.
As we are interested in large graphs, we assume $n \gg 1$.
Let $\langle k \rangle$ and $\langle k^2 \rangle$ be the first and second moment of the node degrees in the original graph $G$, respectively. 
Let $\tau = t/n$ and $n_e(\tau) = E\left[\vert E(\tau n,i) \vert \right]$ be the average number of edges in $S(\tau n,i)$, where $E[\cdot]$ is the expectation operator.
We provide a closed form mean field approximation (Theorem~\ref{lem:ne} in Section~\ref{sec:meanfield}) for $n_e(t)$ when $\tau \ll \langle k\rangle^2/\langle k^2 \rangle$ 
\begin{equation}\label{eq:ne}
  n_e(\tau) = 2 m  \left( 1 - \exp\left({-\mfrac{\langle k^2 \rangle - \langle k\rangle}{\langle k \rangle^2} \tau} \right) \right)  .
\end{equation}
and 
\begin{equation*}\label{eq:ES}
 E[\vert S(\tau n,i) \vert] = \mfrac{n \langle k \rangle}{\langle k^2 \rangle - \langle k\rangle}  \left( 1 - \exp\left({-\mfrac{\langle k^2 \rangle - \langle k\rangle}{\langle k \rangle^2} \tau} \right) \right)  .
\end{equation*}

\begin{figure}[h!!]
\centering
\includegraphics[scale=0.7,type=pdf,ext=.pdf,read=.pdf]{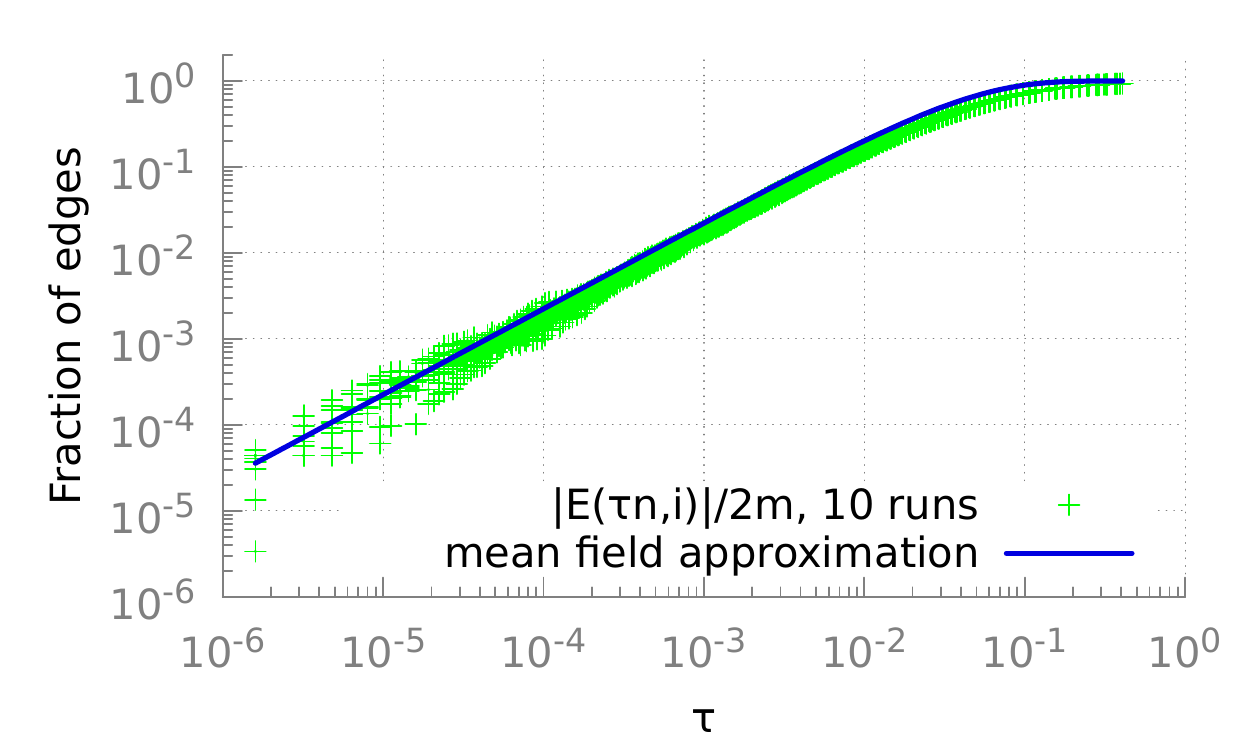}
\caption{{\bf $\bf n_e(\tau)$ in the Gnutella network.} This graph shows the fraction of edges in $S(t,\cdot)$ against $\tau = t/n$ in the Gnutella network. The simulation results of 10 runs are shown in a scatter plot of red crosses.
The mean field approximation in equation~\eqref{eq:ne} is shown as a green line. In the Gnutella network $({\langle k^2 \rangle - \langle k\rangle})/{\langle k \rangle} = 10.6$, which is not as large as in a ``true power law'' network.\label{fig:gnutellane}}
\end{figure}

\begin{figure}[h!!]
\centering
\includegraphics[scale=0.7,type=pdf,ext=.pdf,read=.pdf]{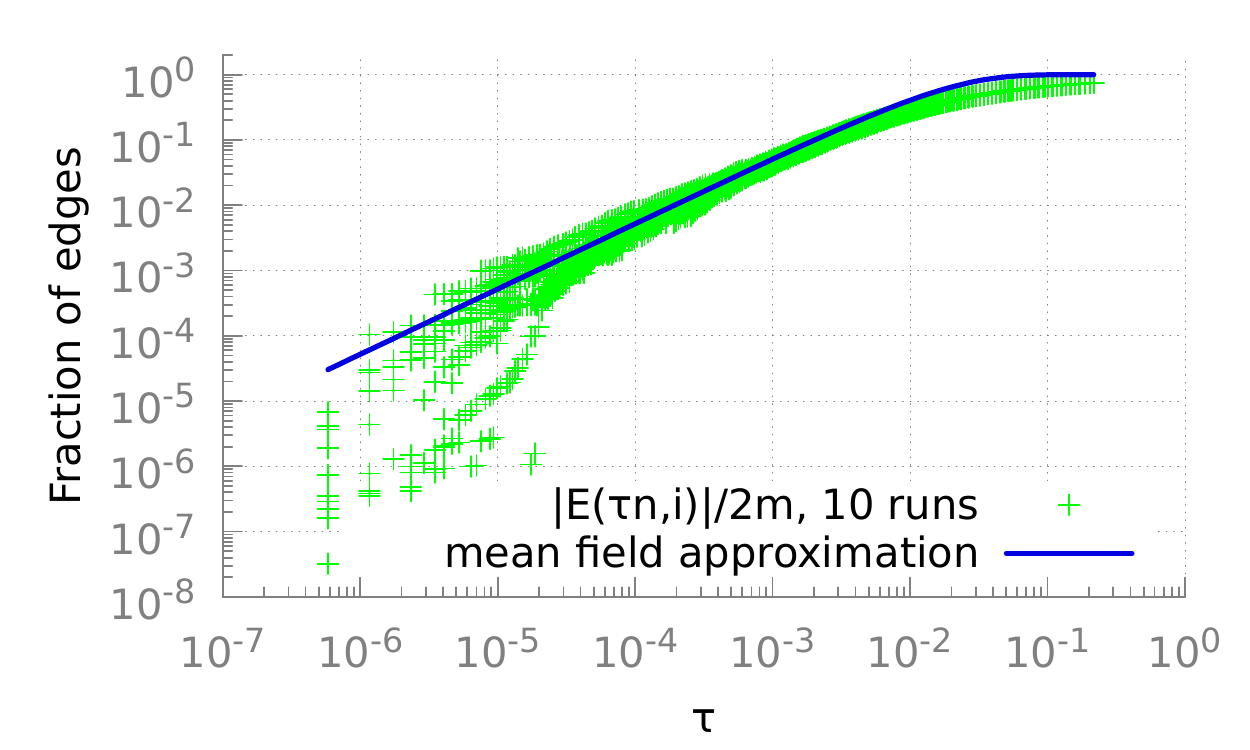}
\caption{{\bf $\bf n_e(\tau)$ in the Flickr network.} Same as Figure~\ref{fig:gnutellane} over the Flickr network. In the Flickr network $({\langle k^2 \rangle - \langle k\rangle})/{\langle k \rangle} = 943.4$.\label{fig:flickrne}}
\end{figure}

Figures~\ref{fig:gnutellane} and~\ref{fig:flickrne} compare our mean field approximation in~\eqref{eq:ne} with simulation results for the Gnutella and Flickr networks.
The Gnutella network cannot be considered a power law network as its maximum degree is quite small (one hundred), but our approximation works for any degree distribution not only for power laws.
Please refer to Table~\ref{tab:traces} for characteristics of these two networks.

To see how $n_e(\tau)$ grows with $B = \beta n$ let's decompose~\eqref{eq:ne} into its Taylor series expansion
\begin{equation}\label{eq:neT}
n_e(\beta) = \mfrac{\langle k^2 \rangle - \langle k\rangle}{\langle k \rangle} \beta n + O(\beta^2) .
\end{equation}
Eq.~\eqref{eq:neT} yields for $\beta \ll 1$
\begin{equation}\label{eq:neTapp}
 n_e(\beta) \approx  \mfrac{\langle k^2 \rangle - \langle k\rangle}{\langle k \rangle} \beta n \, .
\end{equation}
In a power law network in the asymptotic regime $n\to \infty$, i.e., if $k$ is the node degree then it has degree distribution
\begin{equation}\label{eq:pk}
 p_k \propto (\alpha-1) k^{-\alpha}.
\end{equation}
For power law degree distributions the approximation in~\eqref{eq:neTapp} yields
\[
 n_e(\beta) = \begin{cases}
                                       \infty &  \mbox{if }\alpha \leq 3 \\
                                       \beta n \, (\alpha-3)^{-1}  & \mbox{if } \alpha > 3 \, .
                               \end{cases}
\]
While $n_e(\beta) \not \to \infty$ in any real world network, the above equation provides an invaluable hint that $n_e(\beta)$ can be quite large in real life.
Indeed, in the Flickr network we observe
\[
n_e^{\mbox{flickr}}(\beta) \approx 943.4 \times 1,\!700,\!000 \times \beta \, \mbox{ edges} ,
\]
which means that by the end of the crawling the walker collects on average $943.4 \beta n$ distinct edges.
Table~\ref{tab:traces} shows the value of $q=({\langle k^2 \rangle - \langle k\rangle})/{\langle k \rangle}$ for several real world networks.
Before we proceed to the proof of equation~\eqref{eq:ne}, we switch gears and show that under certain conditions a RW starting at $u_i$, $i=1,\dots,h$, visits a node that is also visited by RW $j$, $\forall j \neq i$, with high probability.

\subsection{With high probability two RW sample paths cross.\label{sec:pathsHP}}
In what follows we show that under certain conditions the RW sample paths starting at nodes $u_i$ and $u_j$, $\forall i, j$, cross with high probability.
Assume, without loss of generality, that RW $i$ completes before we start RW $j$.
The intuition behind our result is that for sufficiently large $\vert E(B,i) \vert / 2m$, $S(B,i)$ functions as an attractor for the $j$-th RW, such that with high probability RW $j$ hits the set $S(B,i)$ before $B$ steps.
The above reasoning, however, cannot be applied to a RW that has a long transient; for instance a RW on a graph shaped like dumbbell likely will not satisfy our conditions.

We are interested in the hitting time of RW $j$ into the set $S(B,i)$.
Let $X = \{X_1^{(j)},\dots,X_B^{(j)}\}$ be the set of nodes visited by RW $j$.
Then
\[
  P[ X \cap S(B,i) = \emptyset ]
\]
denotes the probability that RW $j$ and RW $i$ do not meet.
Unfortunately finding the hitting time to a set of nodes in a general graph is a hard problem.
Solutions based on spectral methods (e.g., Chau and Basu~\cite[Theorem 1]{ChauBasu09}) do not yield simple expressions for the general case.
A more detailed discussion is found on Section~\ref{sec:discussion}.
Hence, we consider the following approximation.
Let $B = \beta n$, $n \gg 1$, and
\[
\gamma(B,i) = \frac{\vert E(B,i) \vert}{2 m}
\]
be the fraction of edges that belong to nodes in $S(B,i)$, i.e., nodes visited by the $i$-th RW.
Note that $\gamma(B,i)$ is the steady state probability that RW $j$ visits a node in $S(B,i)$ at any given time step.
As $n\gg 1$ the average value of $\gamma(B,i)$ is
\[
\bar{\gamma}(\beta) = E[\gamma(B ,i) ]  = \frac{n_e(\beta)}{2 m}.
\]
Let $\Delta_n = o(n)$ and assume $\forall w \in \{1,B-\Delta\}$ that for some (possibly small) constant $c > 0$, the following inequality holds
\begin{equation}\label{eq:hit}
P[X_{w+\Delta_n}^{(j)} \in S(\beta n ,i) \vert X_w^{(j)} \not \in S(\beta n ,i)] \geq  c \bar{\gamma}(\beta) .
\end{equation}
That is, if RW $j$ is not at $S(B,i)$ at time $w$ then at time $w + \Delta$ it will be at $S(B,i)$ with probability that is at least a constant factor of $\bar{\gamma}(\beta)$.
In a graph with $o(n)$ mixing time $c \approx 1$ and the inequality in~\eqref{eq:hit} becomes an equality.
However, sublinear mixing time is a sufficient but not necessary condition in~\eqref{eq:hit}, which prompts us to conjecture that~\eqref{eq:hit} holds for a broader class of graphs.
Note that 
\[
  P[ X \cap S(B,i) = \emptyset ] \leq P[ \{X_{1},X_{\Delta+1}, \dots\} \cap S(B,i)  = \emptyset ]
\]
as the probability that RW $j$ at steps $1, \Delta+1, \dots$ does not hit $S(B,i)$ is higher than the probability that RW $j$ at all steps $1,2,\dots,B$ does not hit $S(B,i)$.
Application of \eqref{eq:hit} and the strong Markov property yields
\begin{equation}\label{eq:phit}
P[ X \cap S(B,i) = \emptyset ] \leq \left( 1 - c \bar\gamma(\beta)\right)^{\lfloor \beta n / \Delta \rfloor}.
\end{equation}
From~\eqref{eq:ne}, $c \bar\gamma(\beta) > 0$.
As $n \gg 1$, implies $\lfloor \beta n / \Delta \rfloor \gg 1$
we can put it all together into~\eqref{eq:phit}, yielding, for $\epsilon \ll 1$,
\[
P[ X \cap S(B,i) = \emptyset ] \leq \epsilon.
\]
Thus we conclude that if the condition states in~\eqref{eq:hit} is satisfied then walkers $i$ and $j$ must meet with high probability.
In what follows we provide a proof for~\eqref{eq:ne}.
\subsection{Mean field analysis of the number of edges in $G^\prime(t,i)$.\label{sec:meanfield}}
\begin{theorem}\label{lem:ne}
Let $\tau = t/n  \ll \langle k\rangle^2/\langle k^2 \rangle$ and let $n_e(\tau) = E\left[\vert E(\tau n,i) \vert \right]$ be the average number of edges in $S(\tau n,\cdot)$.
Then
\begin{equation*}
  n_e(\tau) = 2m  \left( 1 - \exp\left({-\mfrac{\langle k^2 \rangle - \langle k\rangle}{\langle k \rangle^2} \tau} \right) \right)  .
\end{equation*}
\end{theorem}
\begin{proof}
By following an edge, a RW reaches a node $v$ with degree $k_v$ with probability proportional to $k_v$.
Node $v$ has average degree higher than the average degree in the graph.
This is a classical case of the inspection paradox and therefore $E[k_v]= \langle k^2 \rangle/\langle k \rangle$.
If $v \not \in S(t,i)$ we are adding at most $k_v - 1$ edges to $E(t,i)$, (the minus one is a consequence of the need to remove the edge the RW followed to reach $v$ which was already in $E(t,i)$).
If $G$ was a random configuration graph, as $t \ll n$ and $n \gg 1$, the probability that any of the $k_v - 1$ edges are already in $E(t,i)$ is negligible.
Thus, by adding $v$ we are adding $q = E[k_v]-1$ edges to $E(t,i)$.

Assuming the RW is sampling edges uniformly at random, we can write the following expression for $n_e$
\begin{align*}
n_e(\tau + 1/n) - n_e(\tau) & = qP[v \not \in S(t,i)] = q\left(1 -  \mfrac{n_e(\tau)}{2m}\right),
\end{align*}
where $n_e(\tau)/2m$ is the probability that a randomly sampled edge is in $E(t,i)$.
Multiplying both sides of the above equation by $n$ and letting $n \to \infty$ yields
\begin{align*}
 \mfrac{n_e(\tau + 1/n) - n_e(\tau)}{1/n} & = nq \left(1 -  \mfrac{n_e(\tau)}{2m}\right) \\
\mfrac{dn_e(\tau)}{d\tau} & = \mfrac{q}{\langle k \rangle} (2m -  n_e(\tau))
\end{align*}
with the boundary condition $n_e(0) = 0$.
The solution to the above differential equation is
\begin{align*}
n_e(\tau) =2 m (1 -  e^{-q\tau/\langle k \rangle}),
\end{align*}
concluding the proof. 
\end{proof}

\begin{lemma}
Let $\tau = t/n \ll \langle k\rangle^2/\langle k^2 \rangle$.
Then
\begin{equation*}
 E[\vert S(\tau n,i) \vert] = \mfrac{n}{\langle k^2 \rangle - \langle k\rangle}  \left( 1 - \exp\left({-\mfrac{\langle k^2 \rangle - \langle k\rangle}{\langle k \rangle^2} \tau} \right) \right)  .
\end{equation*}
\end{lemma}
\begin{proof}
The proof of follows from Lemma~\ref{lem:ne} with  $E[\vert S(\tau n,i) \vert]  = n_e(\tau)/q$.
\end{proof}

\section{The Random Walk Short Path (RWSP) Algorithm} \label{sec:algo}

The RWSP algorithm  is as follows:
Initialize $W(i)=\emptyset$ and $C(i)=\emptyset$, $i=1,\dots,h$.
\begin{enumerate}
\item Start walker $i$ at node $u_i \in U$ with budget $B$, $i=1,\dots,h$.
\item For each walker $i=1,\dots,h$:
\begin{enumerate}
\item At time step $t$ update the ``explored'' graph $G^\prime(t,i)$ as new nodes are visited and new edges are discovered.
\item If the $t$-th visited node, $x_t$, has already been visited by a set of walkers $W \in \{1,\dots,h\}^w$; 
\begin{enumerate}
\item Assign $W(i) \leftarrow W(i) \cup W$; 
\item Assign $C(i) \leftarrow C(i) \cup \{x_t\}$; 
\end{enumerate}
$u_i$ sends a message to $u_j$ advertising that node $u_i$ just found $x_t$ (sending this last message requires tracing back the path $x_t$ to $u_j$ using $j$'s breadcrumbs), $\forall j \in W$.
\item Stop when $t = B$.
\end{enumerate}
\item At the end of walk send $G^\prime(B,i)$ to $u_j$, $\forall j \in W(i)$,  via the nodes in $C(i)$ that have been visited by walker $j$.
\item Upon receiving $(j,v)$, $v \in V$, from walker $j \neq i$: assign $W(i) \leftarrow W(i) \cup \{j\}$ and $C(i) \leftarrow C(i) \cup \{v\}$.
\item Let $G^\star(i) = \bigcup_{r \in W(i)\cup\{i\}} G^\prime(B,j)$ denote the union of all subgraphs known to $u_i$. Compute the minimum spanning tree $T(i)$ on $G^\star(i)$ starting from $u_i$. 
\end{enumerate}
Node $u_i$ uses $T(i)$ to route its messages to nodes $u_j$ on $G$, $\forall j \in W(i)$.
Figure~\ref{fig:progress} shows a snapshot of the algorithm.
In what follows we present our simulation results.

\section{Results} \label{sec:results}

\begin{figure*}[tp!!]
\centering
\subfloat[LiveJournal ($h=4$, $B=0.025\, n$)]{
\includegraphics[width=3.5in,height=2.7in,type=pdf,ext=.pdf,read=.pdf]{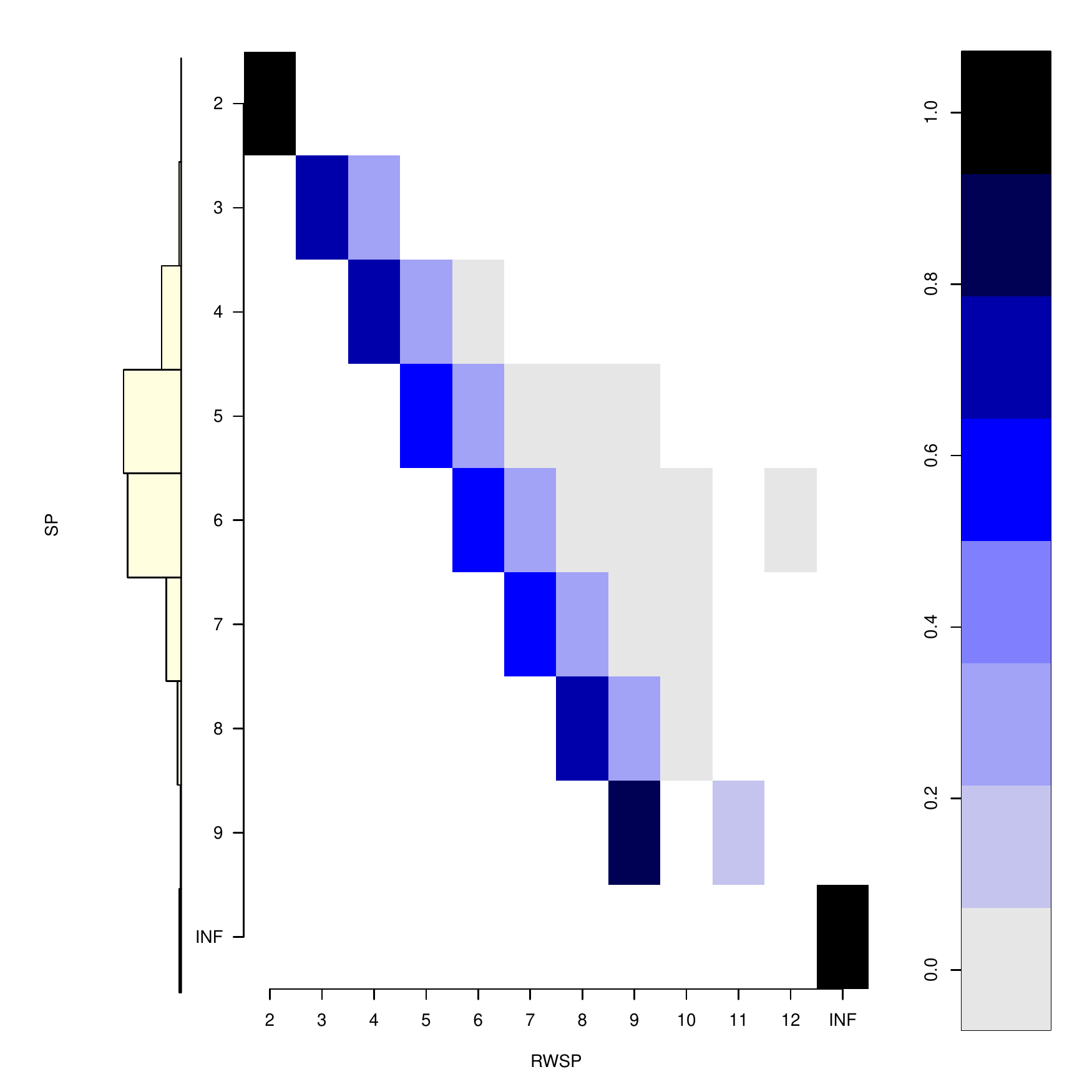}
\label{fig:LJ}
}
\subfloat[Flickr ($h=4$, $B=0.025\, n$)]{
\includegraphics[width=3.5in,height=2.7in,type=pdf,ext=.pdf,read=.pdf]{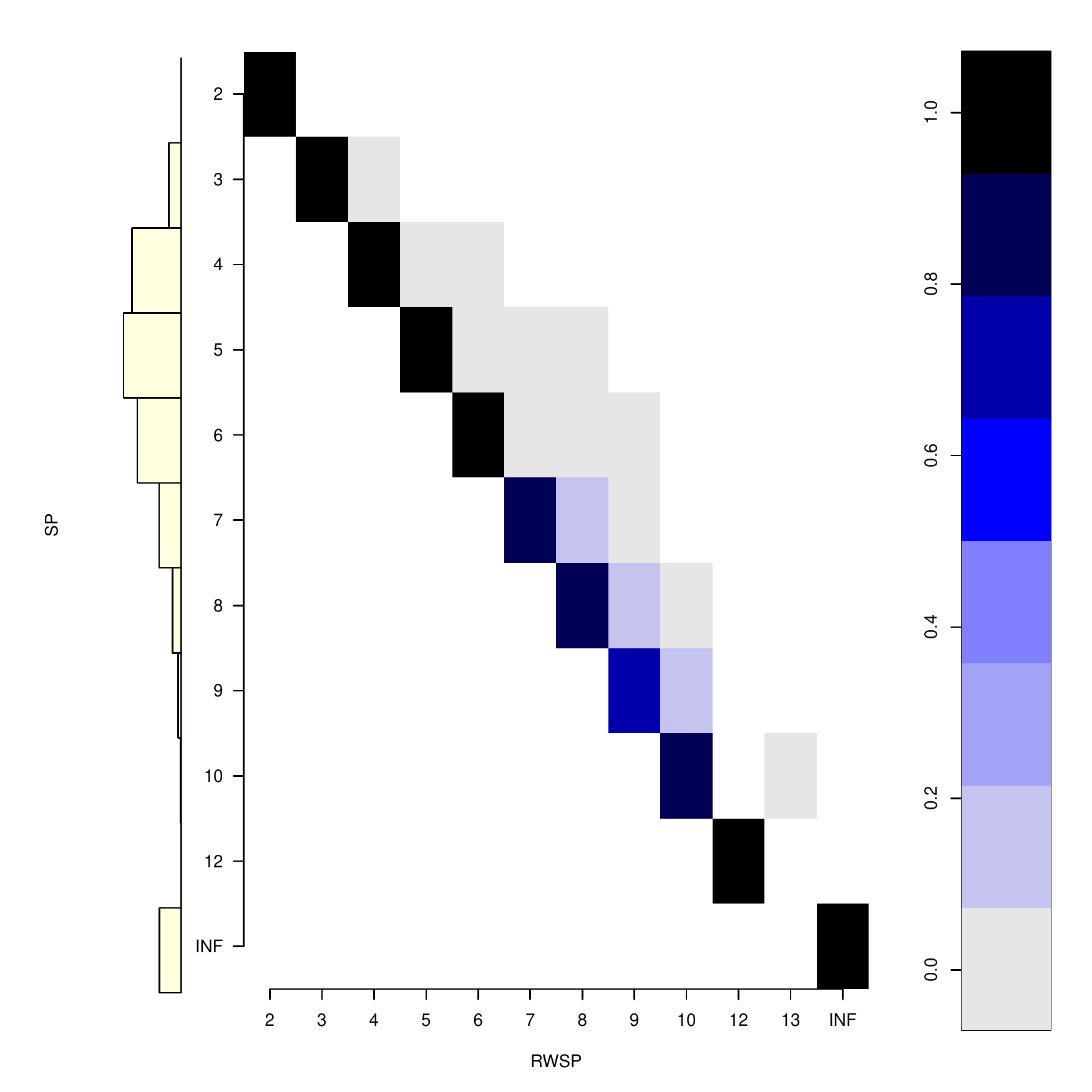}
\label{fig:flickr}
} 
\\
\subfloat[Gnutella ($h=4$, $B=0.05\, n$)]{
\includegraphics[width=3.5in,height=2.7in,type=pdf,ext=.pdf,read=.pdf]{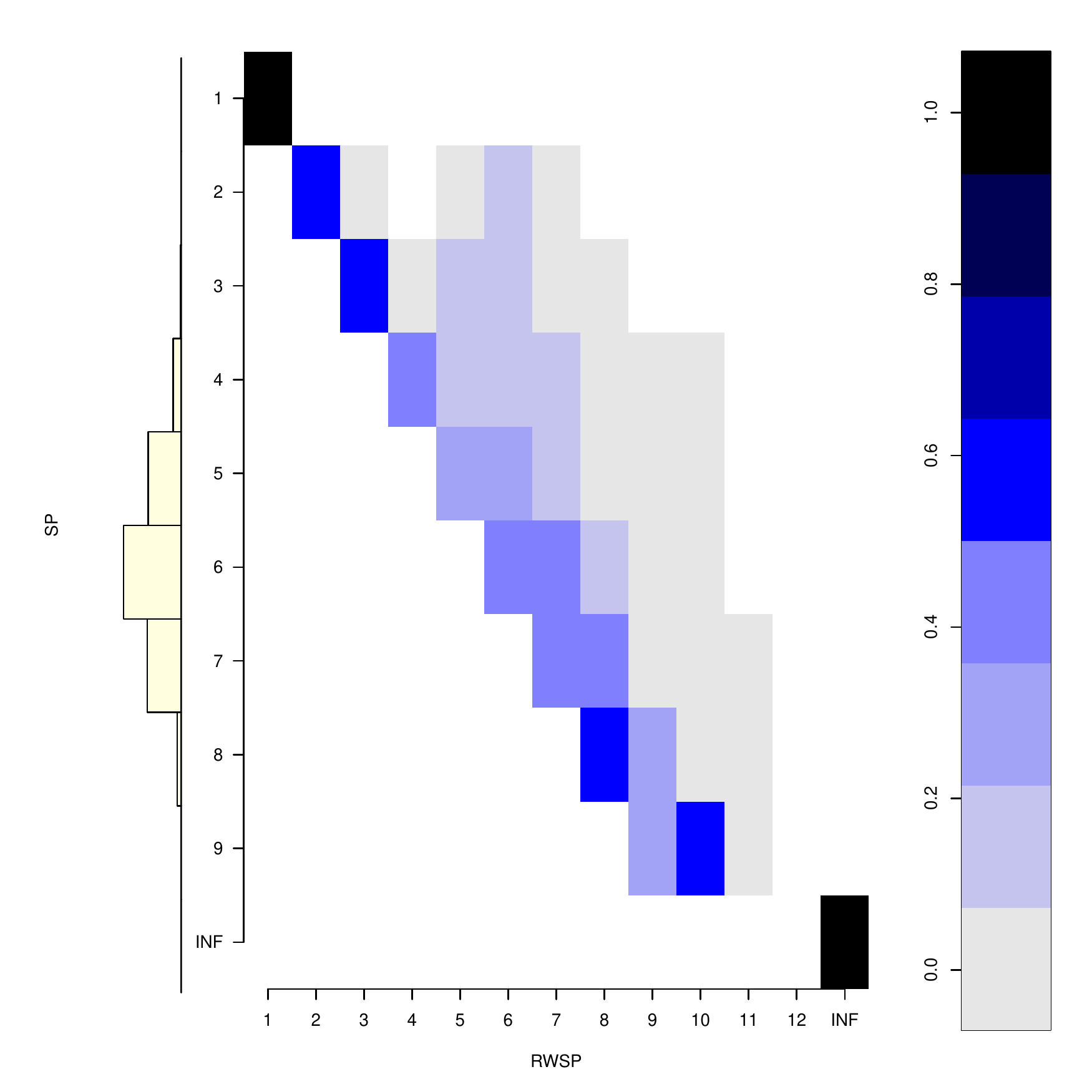}
\label{fig:GNU}
}
\subfloat[AS graph ($h=4$, $B=0.0125\, n$)] {
\includegraphics[width=3.5in,height=2.7in,type=pdf,ext=.pdf,read=.pdf]{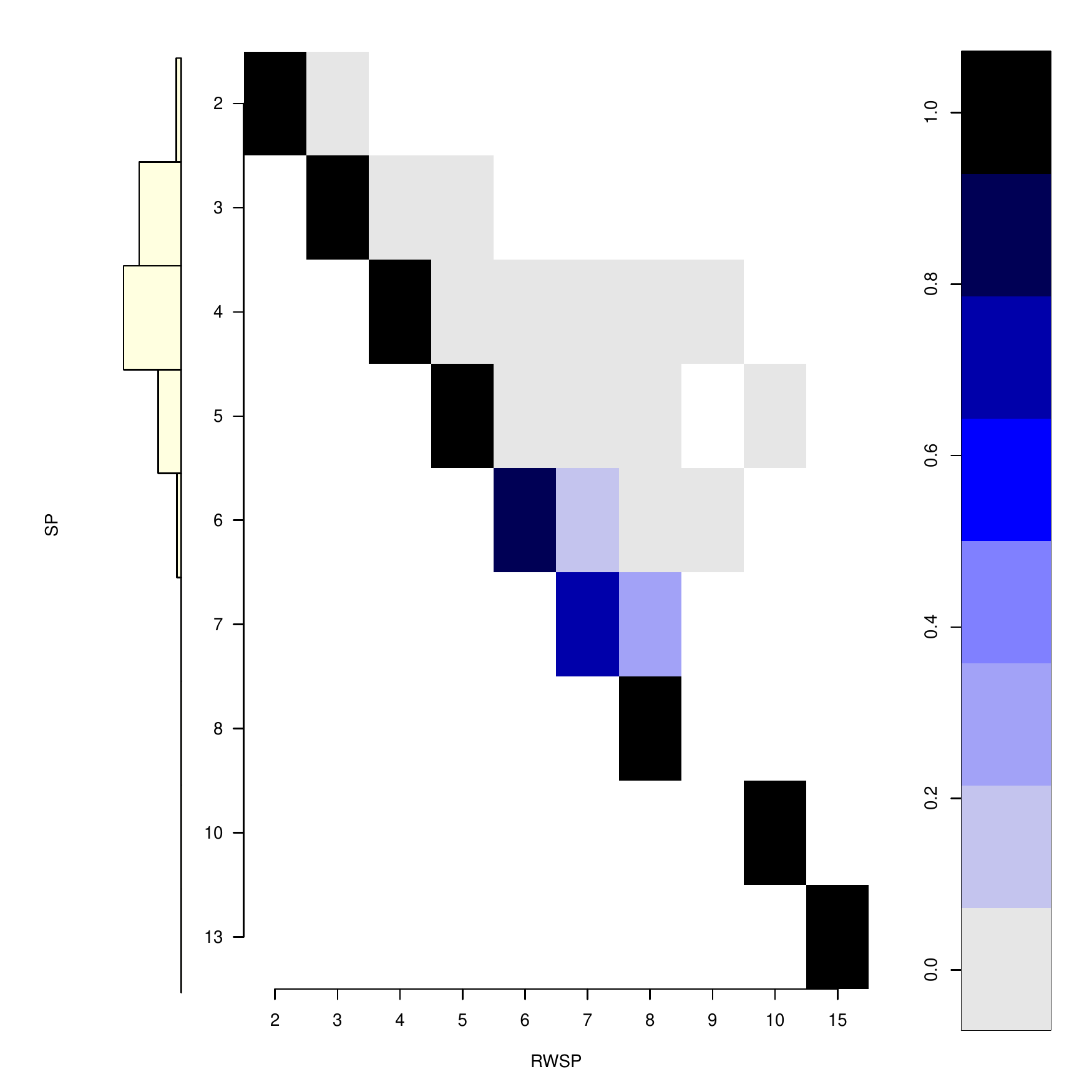}
\label{fig:AS}
}
\\
\subfloat[Enron ($h=16$, $B=0.0125\, n$)] {
\includegraphics[width=3.5in,height=2.7in,type=pdf,ext=.pdf,read=.pdf]{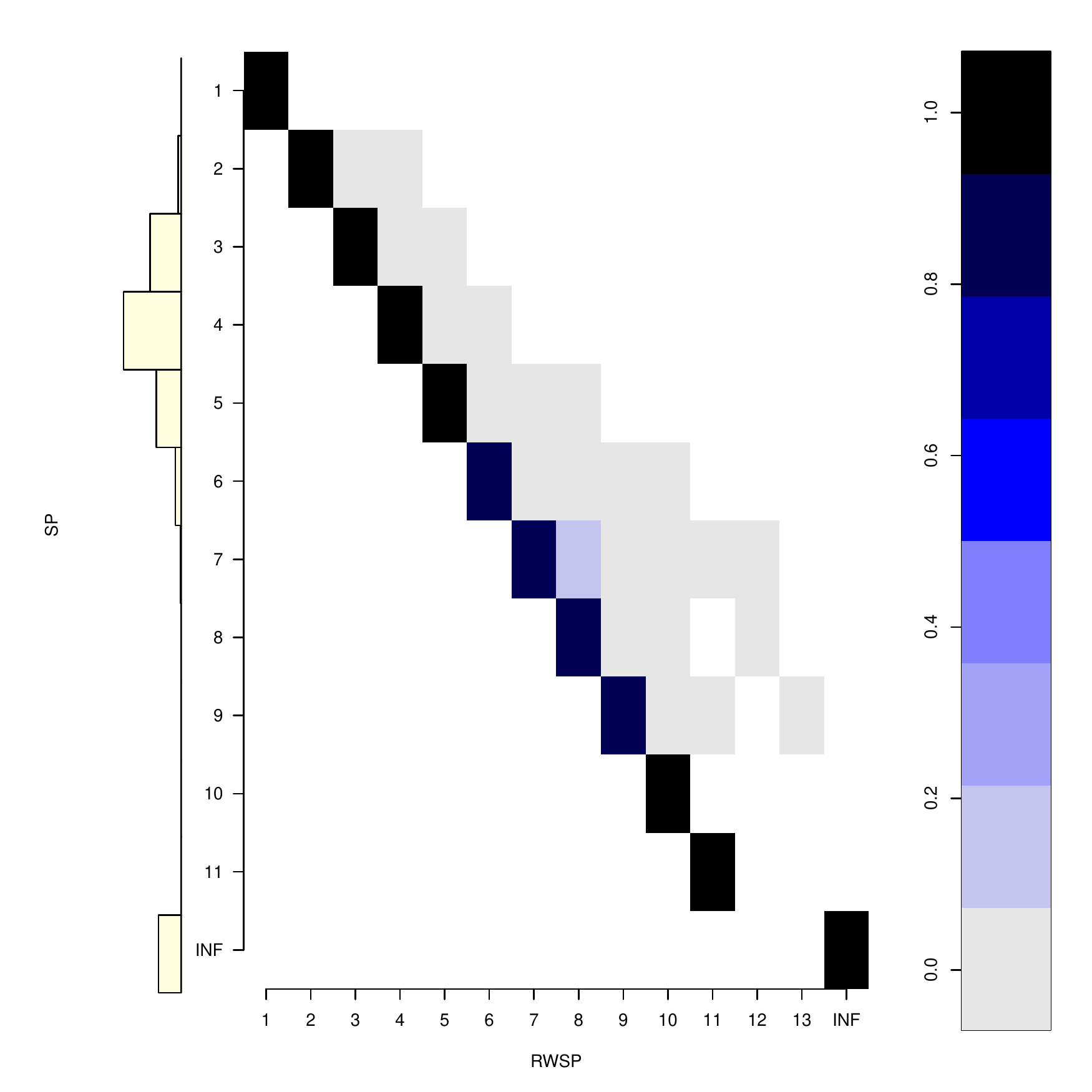}
\label{fig:Enron}
}
\subfloat[Power Grid ($h=4$, $B=0.05\, n$)] {
\includegraphics[width=3.5in,height=2.7in,type=pdf,ext=.pdf,read=.pdf]{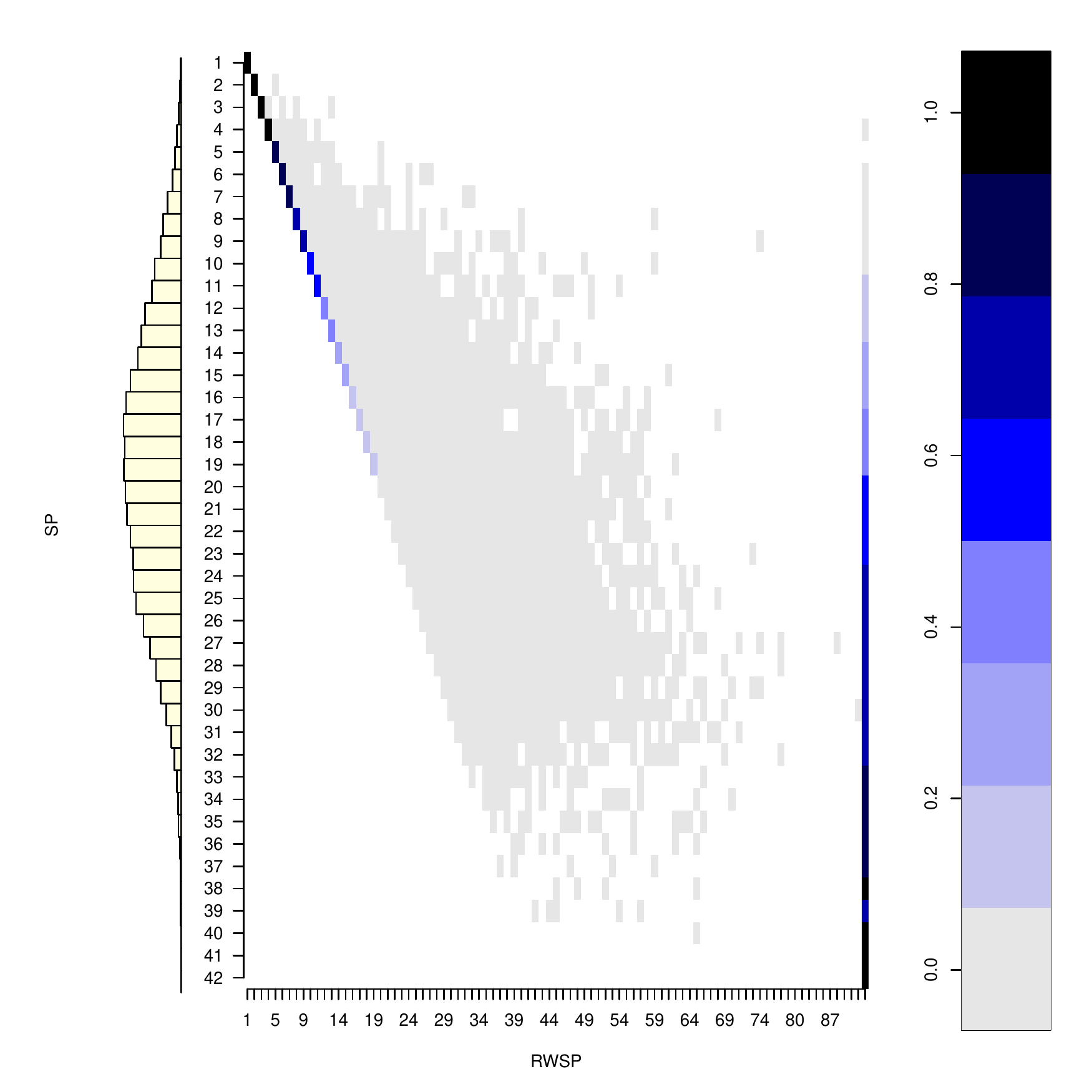}
\label{fig:POWER}
}
\caption{{\bf RWSP vs.\ Shortest Paths.} $h$ walkers and sampling budget $B$. Fraction of RW shortest path lengths (X axis) between nodes vs.\ true shortest path lengths (Y axis) [blue/gray matrix]. Yellow bars next to Y axis show true shortest path lengths distribution. The axis marking INF refers the fraction of times that a node could not reach another node (due to disconnected components). \label{fig:tst}}
\end{figure*}

This section presents our results when simulating RWSP on real world networks.
Table~\ref{tab:traces} summarizes the datasets used in our simulations.
Figure~\ref{fig:LJ} show the fraction of RW shortest path lengths on $G^\star$ defined in Section~\ref{sec:algo} (X axis) between $h=4$ nodes vs.\ the true shortest path lengths (Y axis) [blue/gray matrix]. 
We choose $h=4$ walkers because as we increase $k$ our results get better if $B$ is kept constant.
We also evaluate different values of $h$, $h \in \{2,6,8,16\}$, rescaling $B$ such that for a constant $\beta > 0$, $B = \beta n /h$.
The results for different values of $h$ are similar to those presented here, thus due to space constraints we omit them.

In Figure~\ref{fig:LJ} the yellow bars next to the Y axis show the true all pairs shortest path length distribution. For instance, we in LiveJournal's most of the shortest paths are between 5 and 6 hops. 
The RW sampling budget is $B=2.5\%$ of the nodes in the graph.
A total of $10,000$ runs were used to produce the averages seen in the graph. 
The axis marking INF refers to the fraction of times that a node could not reach another node (due to disconnected components).
In all runs the RWSP was able to find nodes in the same component (we condition the RWSP to start in the giant component).
We also tested RWSP over other networks with similar parameters.

% \begin{figure}[h]
% \centering
% \includegraphics[scale=0.4,type=pdf,ext=.pdf,read=.pdf]{figs/shortest_path_flickr_0.1sampled.log}
% %
% \caption{{\bf RWSP on Flickr social network.} Same as graph in Figure~\ref{fig:LJ} now with the Flickr network. Sampling budget $B=0.025\, n$ of nodes. \label{fig:flickr}}
% \end{figure}

Our results are very promising. 
For Livejournal (Figure~\ref{fig:LJ}) more than $60\%$ of the RWSP paths of any given length are the shortest paths.
Moreover, more than $90\%$ of the RWSP paths of any given length are within one hop of the shortest paths.

We also test RWSP on Flickr, AS graph, a snapshot of the Gnutella network, Enron email dataset with similar results.
On Flickr (Figure~\ref{fig:flickr}) ($B = 0.0125\, n$) more than $85\%$ of the RWSP paths of any given length are the shortest paths; also, all paths are within $2$ hops of the shortest paths.
On Gnutella (Figure~\ref{fig:GNU}) ($B=0.0125\, n$) network we observe more than $60\%$ of RWSP paths of all lengths to be within $2$ hops of the shortest paths.
For the AS graph (Figure~\ref{fig:AS}) (RWSP with $B=0.05 \, n$), more than $65\%$ of RWSP paths of all lengths are the shortest paths; the majority ($>95\%$) of RWSP paths now counting over all lengths are the shortest paths.
On Enron (Figure~\ref{fig:Enron}) ($B=0.05\, n$) we have a result as good as the one for the AS graph.

Figure~\ref{fig:POWER} shows that the only graph in which the RWSP does not perform well, the Power Grid network.
The Power Grid network is not a power law graph and has the largest diameter of any of the previous networks, thus, we expected RWSP to perform poorly.
Consulting Table~\ref{tab:traces} we observe that 
$(\langle k^2 \rangle - \langle k\rangle)/(\langle k \rangle) = 2.9$ is small and thus $G^\prime$ spans only a small fraction of the original graph, making the task of finding a long path close to impossible.

\section{Discussion \& Related Work} \label{sec:discussion}

%ADD discussion about~\cite{AntNet} ant colony optimization. Related to loop-erased random walks.

The problem of topology discovery has received significant attention
in the (communications) networking literature. Broadly, there are two
types of topology discovery -- {\em proactive} and {\em on-demand}. In
proactive ``link-state'' routing schemes such as OLSR~\cite{CJLMQV01},
each node {\em floods} the state of its neighborhood to the entire
network. 
 Since each flood costs $O(m)$ forwardings, where a network
 has $n$ nodes and $m=O(n^2)$ edges, the total overhead of flooding the
 network is $O(m n)$ which is $O(n^3)$ for dense networks. 
Variants of link-state routing such as HSLS~\cite{HSLS} reduce this overhead in
dynamic networks to $O(n^{5/2})$ (amortized over time) by restricting
 the frequency of flooding the link-state updates to nodes located
 farther away; or they optimize the flooding proceeding by using Multi Point Relays~\cite{QVL01}. 
These schemes are advantageous when the
likely traffic patterns are {\em all-to-all}, i.e., every node is
expected to transmit; hence, the cost of topology updates is amortized
over all the nodes.

However, often only a small number $h$ of nodes in the network need to
transmit or receive messages to/from each other. In these situations,
on-demand routing schemes are preferred since they attempt to gather
topology state when they need to communicate~\cite{DSR,AODV}. This is
typically achieved by means of flooding the network with a ``route
request'' message.
% (the ID of the destination is contained in the
% message header). 
When the intended recipient receives this message, it
replies with a ``route response'' message along the reverse path to
the source (this is unicast). 
% If there are $k$ nodes out of $n$ that
% intend to find short paths to each of the other $k-1$ nodes, they will
% initiate such floods. 
In the worst case, this takes $h O(m)$ forwardings in dense networks.
%after ignoring the unicast
%overhead which is $O(k D)$, where $D$ is the diameter of the network.

Opportunistic routing schemes, in contrast, do not attempt to discover the topology but rather make routing decisions adaptively based on the actual transmission outcomes.
The back-pressure algorithm pioneered by Tassiulas and Ephremides~\cite{Tassiulas92} explores and exploits all feasible paths between each source and destination to achieve throughput-optimal routes.
Ant colony optimization is a meta-heuristic technique that uses artificial ants (specially marked packets) to build pheromone trails in the network~\cite{AntNet}.

Random walks have also been used to deliver messages but suffer from excessive delays due to long hitting times~\cite{Servetto02,Mabrouki07}.
Hitting times of random walks have been well studied. For a large
$K\times K$ torus, as $K\rightarrow\infty$, the maximum hitting time
between two nodes is $H_{max}=\Theta(K^2 \log K)$. We showed in our
previous work~\cite{CB11} that for network topologies with node
degrees concentrated around the mean $r$, $H_{max}\approx \frac{n\log
  n}{r^2}$. A corollary to the above is applicable to random geometric
networks at the critical radius~\cite{SR06spec}; in that regime,
$r\approx \log n$, and therefore, $H_{max} = O(n)$. Although in the
worst case hitting times could have a $O(n^3)$ scaling
law~\cite{Lovasz93} (for ``dumbbell'' shaped topologies), in most
commonly occurring topologies, the scaling properties of $H_{max}$
reside between $O(n)$ and $O(n^2)$. Moreover, the hitting time to a
{\em subgraph} drops sharply when the size of the subgraph increases
in comparison to the hitting time to a specific
node~\cite{BG10}. Thus, this could yield lower overhead than on demand
protocols described above, albeit at the cost of a higher routing
stretch (the on demand routing protocols return optimal shortest
paths).

\section{Acknoledgements}
This research was partially funded by U.S. Army Research Laboratory under Cooperative Agreement Number W911NF-09-2-0053 and NSF grant CNS-1065133.
The views and conclusions contained in this document are those of the authors and do not represent the official policies, either expressed or implied, of the U.S. Army Research Laboratory or the U.S. Government. The U.S Government is authorized to reproduce and distribute reprints for Government purposes notwithstanding any copyright notation hereon.

%\balance
%\small
\bibliographystyle{plain}
%\bibliography{epidemic_bibliography}
\bibliography{references}

\end{document}